# Photoluminescence from $Bi_5(GaCl_4)_3$ molecular crystal

Hong-Tao Sun,*[ab] Beibei Xu,[c] Tetsu Yonezawa,[a] Yoshio Sakka,[d] Naoto Shirahata,[e,f] Minoru Fujii,[g]
5  Jianrong Qiu,[c] and Hong Gao [b]

$Bi_5(GaCl_4)_3$ sample has been synthesized through the oxidation of Bi metal by gallium chloride ($GaCl_3$) salt. Powder X-ray diffraction as well as micro-Raman scattering results revealed that, in addition to crystalline $Bi_5(GaCl_4)_3$ in the product, amorphous phase containing $[GaCl_4]^-$ and $[Ga_2Cl_7]^-$ units also exist.
10 The thorough comparison of steady-state and time-resolved photoluminescent behaviors between $Bi_5(GaCl_4)_3$ product and $Bi_5(AlCl_4)_3$ crystal leads us to conclude that $Bi_5^{3+}$ is the dominant emitter in the product, which gives rise to the ultrabroad emission ranging from 1 to 2.7 μm. Detailed quantum chemistry calculation helps us assign the observed excitations to some electronic transitions of $Bi_5^{3+}$ polycation, especially at shorter wavelengths. It is believed that our work shown here not only is helpful
15 to solve the confusions on the luminescent origin of bismuth in other material systems, but also serves to develop novel broadband tunable laser materials.

# Introduction

Bismuth is one of the most thoroughly investigated main group elements, which exists in a wide array of functional materials such as magnets, superconductors, thermoelectric and spintronic materials.[1-4] Besides this, bismuth doped materials show smart optical properties, arising from its diverse oxidation states and profound propensity to form clusters,[5,6] among which the most stable form is +3. In recent years, particular attention has been paid on ultrabroad near infrared (NIR) emitting materials containing bismuth, thanks to their great potentials for fibre lasers,[8] optical amplifiers (ref.9a) and bioimaging (ref.10). Untill now, NIR photoluminescence (PL) has been observed in glasses,[7-10] conventional crystals,[11-15] ionic liquids,[16] and molecular crystals (ref.17). However, several important challenges and questions on the NIR PL mechanism of Bi remain, despite the extensive amount of research on this topic.[7-19] An important practical question is whether Bi with an identical oxidation state (i.e., the same type of Bi active center) contributes to the NIR emission in the aforementioned materials. Although rather conflicting results exist in the literature, recent work clearly revealed that subvalent Bi (i.e., the average valence state of Bi is +1 or between + 1 and 0) is the active center in some peculiar systems.[16-18] However, for most materials containing Bi, it is still a hard task to finally determine the detailed information of subvalent Bi, and it is rather inconvincing if we simply assign the NIR PL in different systems to the same type of Bi active center, considering the large difference of steady-state and time-resolved PL and of absorption spectra. Indeed, the reports of the synthesis and photophysical properties of subvalent bismuth date back to half a century ago,[5a-5e] but it was not until recently that luminescent properties of these species were studied.[16-19] The complex structural characteristics of subvalent Bi species (ref. 5 and 6) imply that thorough characterization of not only photophysical properties, but also of structural features is prerequisite to know the exact PL mechanism in Bi containing systems.[12c,17] A second practical issue is that multi-type rather than single-type Bi active center may exist in some materials. For instance, Sun *et al.* reported that $Bi_5^{3+}$ and $Bi^+$ emitters can be stabilized by a Lewis acidic liquid, which show ultrabroad NIR PL with a lifetime of around 1 μs.[16] It is reasonable to expect that these active centers may have chances to exist in other materials such as glasses and crystals. A final unresolved critical issue involves the unknown local coordination environments of Bi in most materials. This is a great obstacle to the understanding of Bi-related NIR emission behaviors, since most researchers merely reported the PL-related results and did not show definitely convincing results related to structural analysis, owing to the limited characterization techniques available for those systems. Given these issues, it is believed that finding new NIR-emitting Bi related materials with simple and well-known structures may provide a much clearer picture on the photophysical properties of Bi, which could serve to solve above confusions.

Recently, Sun *et al.* demonstrated that $Bi_5(AlCl_4)_3$ microsized crystal exhibits extremely broad NIR PL with a full width at the half maximum (FWHM) of > 510 nm.[17a] This material has a clear crystal structure, that is, the local coordination environments of Bi, Al, and Cl are well defined using X-ray diffraction and Raman scattering techniques, which result in a convincing interpretation of the NIR PL behaviors. However, there still remain the following scientific issues to be addressed. The first is that some dissociative subvalent bismuth such as $Bi_5^{3+}$ and $Bi^+$ might be left on the surface of the final product, given that these centers exist in the mother liquid,[16] although it proves to be easier to separate the product from the reactants in comparison to molten salt approach.[5,6,17,18] It is also possible that the defects in the crystals partially contribute to this emission, since a room-temperature method was adopted to synthesize it. Subsequent experiments revealed that even after four weeks, the red powders continue to separate out from the mother liquid. Thus, it is possible that the luminescent reaction intermediate existing in the final product (i.e., $Bi_8^{2+}$),[17b] could affect the emission lineshape. Further extension of the reaction time will, in a sense, increase the quality of the obtained crystals. The second issue is that, in that contribution, Sun *et al.* explained the single-photon emission phenomena using linear combination of atomic orbitals (LCAO)-molecular orbital (MO) theory.[17a, 20] Although this method was proved to be fortuitously good to deal with the electronic excitation process of bismuth polycations,[20] it is believed that carrying out a more detailed theoretical calculation using state-of-the-art quantum-chemistry softwares available may be helpful to gain more luminous clues or conclusions related to its photophysical properties.

Herein, we present the experiemental and theoretical studies of the PL behaviors of $Bi_5(GaCl_4)_3$ crystal prepared through the oxidation of Bi metal by gallium chloride salt. The product shows superbroad emission ranging from 1 to 2.7 μm with some intrinsic excitation bands in the visible and NIR spectral regions. In view of the structural similarity of $Bi_5^{3+}$ units in $Bi_5(GaCl_4)_3$ and $Bi_5(AlCl_4)_3$ crystals, we also compared the observed excitation/emission behaviors of $Bi_5(GaCl_4)_3$ with $Bi_5(AlCl_4)_3$, which helps us obtain a deeper understanding of the PL properties of $Bi_5^{3+}$. Furthermore, we carried out a detailed quantum chemistry calculation on $Bi_5^{3+}$ polycations, which leads us to attribute some observed excitation bands to specific electronic transitions of $Bi_5^{3+}$ unit, especially in the UV-Vis spectral ranges.

## Experimental details

### Materials synthesis

Anhydrous $GaCl_3$ (Sigma-Aldrich, 99.999%) and Bi (Alfa Aesar, 99.999%) were used without further purification. Because of the high moisture sensitivity of the anhydrous $GaCl_3$, all handling of the chemicals and sample preparations were performed under dry $N_2$ atmosphere in a glovebox (< 2 ppm $H_2O$; < 0.1 ppm $O_2$). All glassware was dried at 400 °C prior to use. The $Bi_5(GaCl_4)_3$ crystals were synthesized through the oxidation of Bi metal by gallium chloride salt.[6a] $GaCl_3$ and Bi were mixed at a molar ratio of 4:1, and then sealed in an ampule under vacuum. The reaction mixture was equilibrated at 170 °C for 48 h before being slowly cooled to 80 °C at a speed of 1 °C/h. It was observed that the slow cooling of the mixture results in a gradual formation of a single liquid phase and a concomitant precipitation of well shaped rhombohedra1 crystals with metallic luster. The product formed in the bottom of the ampule is black in bulk but brick red when powdered, quite similar to microsized $Bi_5(AlCl_4)_3$ powders. It is necessary to note that the slow cooling of the mixture and using high-purity Bi metal are rather important for obtaining large $Bi_5(GaCl_4)_3$ crystal, although amorphous black phases were found to intersperse in the cystal. It was also found that some colorless powders were attached on the inner wall of the ampule, which is believed to be the left (unreacted) $GaCl_3$. To evaluate the difference of PL behaviors between $Bi_5(GaCl_4)_3$ and $Bi_5(AlCl_4)_3$, the latter was synthesized according to the procedure as reported in ref.17a, and $Bi_5(AlCl_4)_3$ was seperated from the mother liquid after six months. The obtained powders were transferred to bottles or capillaries for the following powder X-ray diffraction (PXRD), Raman, steady-state and time-resolved PL spectroscopic measurements.

**Materials characterization**

The products were characterized by X-ray diffractometer (Rigaku-RINT Ultima3, λ=1.54056 Å), which were grounded into powders, sealed in 1 mm Hilgenberg borosilicate capillaries and kept spinning during the measurement. Raman spectrum was performed by using a Jobin Yvon T-64000 system under the excitation of 514.5 nm line from an $Ar^+$ laser. To avoid the decomposition of $Bi_5(GaCl_4)_3$ by focused laser beam, Raman spectrum should be taken at a low excitation power and in a short exposure time, thus resulting in a relatively poor signal-to-noise ratio. The excitation-emission matrix (EEM) was taken by a Horiba NanoLog spectrofluorometer equipped with a monochromated Xe lamp and a liquid $N_2$ cooled photomultiplier tube (PMT) (Hamamatsu, R5509-72). The powder was sandwiched between a glass microscope slide and a quartz cover glass and the two glasses were sealed with glue. Emission spectra were taken at different excitation wavelengths from 250 to 910 nm with 20 nm intervals. It is noted that all spectra were corrected for the spectral response of the detection system. Time-resolved PL measurements were performed by detecting the modulated luminescence signal with a PMT (Hamamatsu, R5509-72), and then analyzing the signal with a photon-counting multichannel scaler. The excitation source for the time-resolved PL measurements was 488 nm light (pulse width: 5nsec; frequency: 20Hz) from an optical parametric oscillator pumped by the third harmonic of a Nd:YAG laser. The detailed quantum chemistry calculation on $Bi_5^{3+}$ polycation is shown in the following section.

**Results and discussion**

The orange/red crystalline powder of $Bi_5(GaCl_4)_3$ was characterized by powder X-ray diffraction (PXRD) (Fig. 1). The diffraction pattern corresponds well to that of $Bi_5(GaCl_4)_3$ single crystal (CSD No: 414089).[6h] The $Bi_5(GaCl_4)_3$ crystal is built from $Bi_5^{3+}$ trigonal bipyramids and $[GaCl_4]^-$ units (Fig. 2). It is necessary to note that the the compound reported here is not isostructural with the previously characterized $Bi_5(AlCl_4)_3$;[6f, 6h,17a] $Bi_5(GaCl_4)_3$ and $Bi_5(AlCl_4)_3$ crystallize in the space group 161 and 167, respectively. The bond lengths of Bi1-Bi1, Bi1-Bi2, and Bi1-Bi3 in $Bi_5(GaCl_4)_3$ are 3.306, 3.020, and 2.968 Å, respectively (Fig. 2b), and the average length is 3.098 Å.

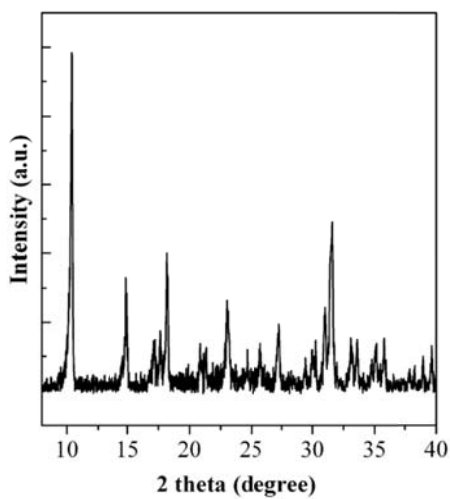

**Fig. 1** PXRD spectrum of $Bi_5(GaCl_4)_3$ after removal of the background.

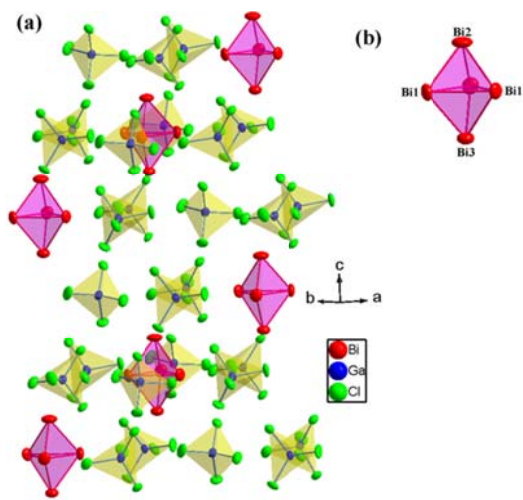

**Fig. 2** (a) The crystal structure of $Bi_5(GaCl_4)_3$ consisting of triagonal bipyramidal $Bi_5^{3+}$ clusters and tetrahedral $[GaCl_4]^-$ units. (b) View of a single $Bi_5^{3+}$ cluster.

It is well known that Raman spectroscopy is a powerful tool to investigate the structure of molecular crystals containing bismuth.[6a,16,17a] Here we employed this technique to check the structure of the obtained product. Although a relatively large crystal can be obtained by the method reported here, the microscopic exmination of the powdered samples found that $Bi_5(GaCl_4)_3$ was mixed with some black amorphous substance. As shown in Fig. 3, the red and black phases demonstrate quite different Raman scattering bands. For the red phase, two notable scattering bands located at 136 and 340 cm$^{-1}$ are observed, which can be assigned to the vibrational modes of $Bi_5^{3+}$ trigonal bipyramids and $[GaCl_4]^-$ units, respectively.[6a,16,17a] Unfortunately, the poor signal-to-noise ratio in the range of < 100 cm$^{-1}$ makes it difficult to do meaningful assignments of vibrational modes. In contrast, the black phase shows a new scattering band at 256 cm$^{-1}$, which can be assigned to the characteristic vibration of $[Ga_2Cl_7]^-$ units.[6a] This becomes understandable if we further consider the formation mechanism of $Bi_5(GaCl_4)_3$ crystal. When the mixture of Bi and $GaCl_3$ was heated at 170 °C, the following reaction may occur,[6a]

$$10Bi + (9n+3)GaCl_3 \rightarrow 2Bi_5^{3+} + 3Ga^+ + 9Ga_nCl_{3n+1}^- \quad (1)$$

When slowly cooling down the reactant, $Bi_5^{3+}$ polycations are stablized by $[GaCl_4]^-$ units and crystalline $Bi_5(GaCl_4)_3$ forms. Some $[Ga_2Cl_7]^-$ also exists in the final product, and may act as the stabilization agent for some unknown charged bismuth species. It is noteworthy that no crystalline impurity was found in the final product except for $Bi_5(GaCl_4)_3$ crystal (Fig. 1), suggesting that the black phase should exist in the amorphous state.

Next, we took EEM of the sample to examine its photophysical behavior. Clearly, the sample demonstrates three noticeable excitation-emission regions at the excitation maximam of 290, 530 and 800 nm (Fig. 4). It is clear that at the excitation wavelenghs of 750-910 nm, the sample tends to show much stronger emission. Unfortunately, owing to the detection limitation of PMT used, signal over 1600 nm can not be recorded. It is interesting to note that the down-converted emission profile is rather broad with an emission maxima over 1600 nm, whose shape is almost identical when excited by different wavelengths (Fig. 5a). Further detailed survey of the emission characteristics of the product revealed that there are two weak excitation regions shouldering/peaking at 390 and 640 nm (Fig. 4 and 5b). These bands are the intrinsic excitation bands of $Bi_5^{3+}$. This assignment is well supported by its absorption spectrum.[6a,17a,20] However, the 290 nm excitation band can not be attributable to the electronic transition of $Bi_5^{3+}$. Previous results clearly revealed that $Bi^{3+}$ ion has an absorption band at around 290 nm corresponding to $^1S_0$-$^3P_1$ transition.[21] The appearance of this band suggests that partial $Bi^0$ was oxidized into $Bi^{3+}$ and embedded in the final product, which could act as an sensitizer of $Bi_5^{3+}$ polycation.

To clarify the emission behavior of $Bi_5^{3+}$ polycations, we also studied the PL spectra of $Bi_5(AlCl_4)_3$, which was synthesized according to the method reported by Sun et al..[17a] It is noted that the red powders were seperated from the mother liquid six months later after mixing of the reactants to ensure the reaction went to completion. As shown in Figs. 6 and 7a, $Bi_5(AlCl_4)_3$ has similar NIR excitation/emission bands with $Bi_5(GaCl_4)_3$. However, there remain the following differences between them. First, $Bi_5(AlCl_4)_3$ does not show 290 nm excitation band (Fig. 6 and 7b), suggesting that no $Bi^{3+}$ exists in $Bi_5(AlCl_4)_3$ sample. Second, $Bi_5(AlCl_4)_3$ sample displays comparable excitation bands in UV-visible and NIR regions (Fig. 7b), while $Bi_5(GaCl_4)_3$ shows a relatively weaker excitation band in UV-visible range (Fig. 5b). This possibly results from their different inherent excitation/emission abilities as a result of the structural difference.[6a,17a,20]

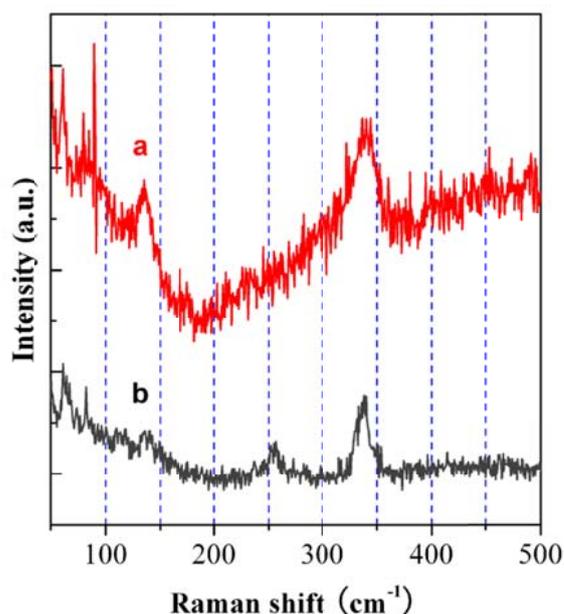

**Fig. 3** Micro-Raman spectra of the $Bi_5(GaCl_4)_3$ product under the excitation of 514.5 nm line from an $Ar^+$ laser. The upper (a) and down (b) spectra were taken from the red and black phases, respectively. Compared with $Bi_5(AlCl_4)_3$, it was found that the sample is very sensitive to the excitation power, and longer exposure time results in destroying the structure. Thus, we took the Raman

spectrum using a low excitation power and in a short time (< 20 s). The sample morphologies were checked carefully after the measurement.

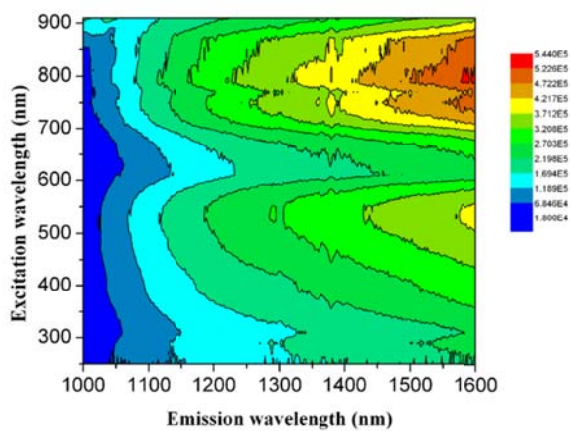

**Fig. 4** Contour EEM plot of the obtained $Bi_5(GaCl_4)_3$ sample. Unfortunately, the high-sensitivity PMT used here does not allow us to take the signal over 1.6μm due to its detection limit.

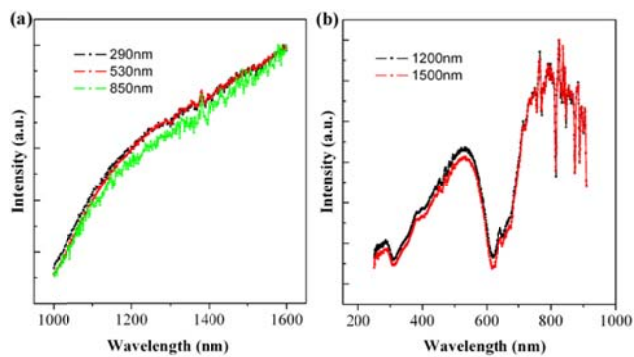

**Fig. 5** (a) PL spectra of the obtained $Bi_5(GaCl_4)_3$ sample under the excitation of 290, 530, and 850 nm. (b) PLE spectra when monitored at 1200 and 1500 nm. Note that the sharp peaks in the range of 750-910 nm result from measurement artefacts.

In order to record the signal over 1.6 μm, we tried to take the PL spectra of both samples by a spectrofluorometer (FLS-920, Edinburgh Instruments) equipped with a InSb photodetector. The excitation source is an 808 nm laser diode. The spectra were corrected for spectral response of the detection system Interestingly, we observed that the $Bi_5(AlCl_4)_3$ sample demonstrates ultrabroad emission band ranging from 1.0 to 2.7 μm, whose peak and FWHM are at 1800 nm and ca. 534 nm, respectively. Although the emission spectrum of $Bi_5(GaCl_4)_3$ sample is very noisy, we could see that its lineshape is similar to that of $Bi_5(AlCl_4)_3$. This suggests that the slight difference of the structures of $Bi_5(GaCl_4)_3$ and $Bi_5(AlCl_4)_3$ crystals does not strongly influence the photophysical properties of $Bi_5^{3+}$ units. On the other hand, the nearly symmetric emission lineshape from the products obtained here, especially for $Bi_5(AlCl_4)_3$ crystal, implies that fewer defects and/or impurities exist in the samples, in comparison with the luminescent $Bi_5(AlCl_4)_3$ crystal reported before.[17a]

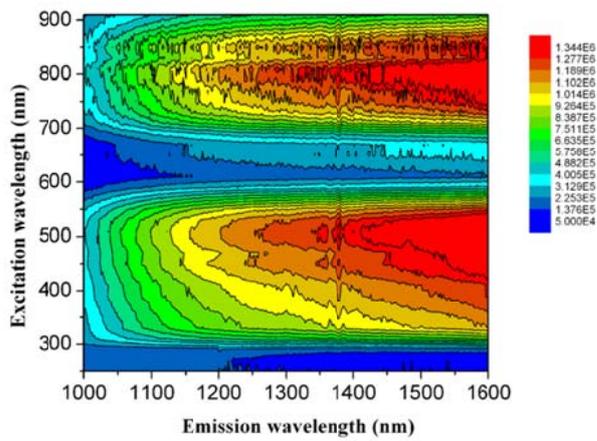

**Fig. 6** Contour EEM plot of $Bi_5(AlCl_4)_3$.

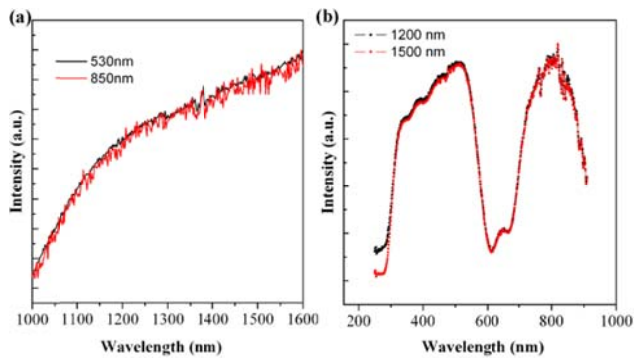

**Fig. 7** (a) PL spectra of $Bi_5(AlCl_4)_3$ under the excitation of 290, 530, and 850 nm. (b) PLE spectra when monitored at 1200 and 1500 nm. Note that smaller slit widths for the excitation and emission were used when taking above spectra, since $Bi_5(AlCl_4)_3$ shows much stronger emission than $Bi_5(GaCl_4)_3$. This results in a relatively flat curve in the range of 750-910 nm in comparison to that shown in Fig.5b.

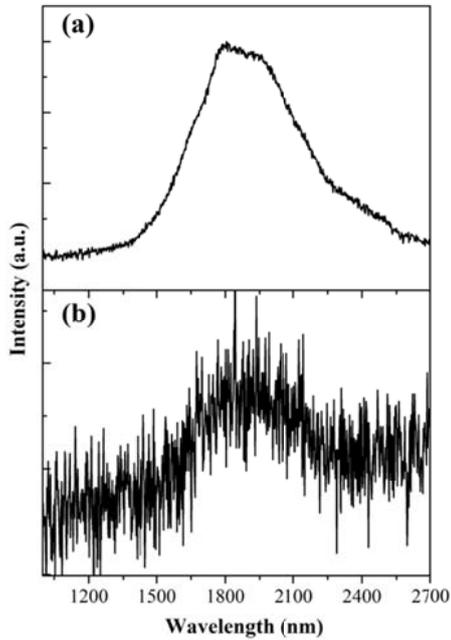

**Fig. 8** PL spectra of $Bi_5(AlCl_4)_3$ (a) and $Bi_5(GaCl_4)_3$ (b) samples under 808 nm excitation.

We next investigated the time-resolved spectra of $Bi_5(GaCl_4)_3$ and $Bi_5(AlCl_4)_3$ samples. As displayed in Fig. 9a, the $Bi_5(GaCl_4)_3$ sample demonstrates nonexponential decays at 1160, 1300 and 1500 nm under the exciation of 488 nm pulsed light. The effective lifetimes at 1160, 1300, and 1500 nm were calculated to be 1.60, 1.67 and 1.79 μs, respectively. In contrast, the decays of $Bi_5(AlCl_4)_3$ sample can be well fitted by double exponentials. It is clear that the fast decay components dominate in the decay curves at all detected wavelengths (Fig. 9b). The most possible reason for the slight deviation of a single-exponential decay for this sample lies in the high concentration of $Bi_5^{3+}$ and rather short spatial distance of adjacent clusters, which results in nonradiative relaxation of $Bi_5^{3+}$ from the electronically excited states. Although $Bi_5(GaCl_4)_3$ sample was synthesized by a "high" temperature approach, the complex multi-exponential decays suggests that there might be structural defects in the product, which introduce nonradiative decay pathways for the deexcitation of $Bi_5^{3+}$.

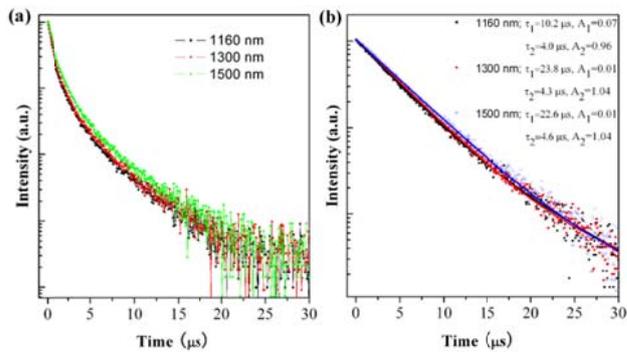

**Fig. 9** Time-resolved PL spectra monitored at 1160, 1300 and 1500 nm for (a) $Bi_5(GaCl_4)_3$ and (b) $Bi_5(AlCl_4)_3$ samples under the excitation of 488 nm pulsed light. The temporal resolution is as high as 80 ns. The decay curves in (b) were fitted by double exponentials.

In the next step, to obtain a clearer picture on the photophysical property of $Bi_5^{3+}$ polycation, we performed detailed quantum chemistry calculations using the Amsterdam Density Functional (ADF) program package developed by Baerends *et al.*.[22] The geometries of $Bi_5^{3+}$ *obtained from the XRD analyses*, as shown in Fig. 10, were used for the following calculations.[6f,6h] Spin-restricted density functional theory (DFT) was employed to determine energies and compositions of excited states of $Bi_5^{3+}$ polycation, using the Hartree-Fock method. The Slater type all-electron basis set utilized in the DFT calculations is of triple-zeta polarized (TZP) quality. The allowed and forbidden electronic transitions were calculated using a Davidson method. The ADF numerical integration parameter was set to 3.0 in all calculations. Scalar relativistic effect was taken into account for DFT calculations.

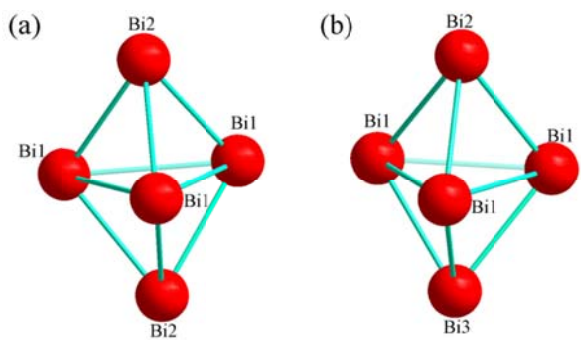

**Fig. 10** (a) $Bi_5^{3+}$ polycation in $Bi_5(AlCl_4)_3$. Bond length (Å): Bi1-Bi1 3.3265(17) and Bi1-Bi2 3.0124(11). (b) $Bi_5^{3+}$ polycation in $Bi_5(GaCl_4)_3$. Bond length (Å): Bi1-Bi1 3.3062(14), Bi1-Bi2 3.020(4), and Bi1-Bi3 2.968(4).

**Table 1.** The calculated electronic transitions of $Bi_5^{3+}$ polycation as shown in Fig. 10a. S-T and S-S represent the singlet-singlet and singlet-triplet transitions, respectively. Note that when the transition probability is very small, the oscillator strength is regarded as 0, in both cases of S-T and S-S transitions.

| No | Wavelength (nm) | Oscillator strength | Symmetry | type |
|---|---|---|---|---|
| 1 | 1415.4 | 0 | $A_1'$ | S-T |
| 2 | 975.5 | 0 | $E'$ | S-T |
| 3 | 778.4 | 0 | $A_2'$ | S-T |
| 4 | 631.5 | 0 | $E''$ | S-T |
| 5 | 567.7 | 0 | $A_2''$ | S-T |
| 6 | 485.2 | $2.20 \times 10^{-33}$ | $A_2'$ | S-S |
| 7 | 483.6 | 0 | $E'$ | S-T |
| 8 | 479.9 | 0 | $E''$ | S-T |

| No | Wavelength (nm) | Oscillator strength | Symmetry | type |
|---|---|---|---|---|
| 9 | 479.4 | 0 | $A_1''$ | S-T |
| 10 | 448.1 | 0 | $A_1''$ | S-S |
| 11 | 447.5 | $3.81 \times 10^{-3}$ | $E'$ | S-S |
| 12 | 434.0 | 0 | $E''$ | S-T |
| 13 | 396.1 | $1.17 \times 10^{-34}$ | $A_1'$ | S-S |
| 14 | 370.1 | 0 | $E''$ | S-T |
| 15 | 361.0 | 0 | $E'$ | S-T |
| 16 | 350.0 | 0 | $A_1'$ | S-T |
| 17 | 322.7 | $8.94 \times 10^{-3}$ | $E'$ | S-S |
| 18 | 309.0 | $1.86 \times 10^{-3}$ | $A_2''$ | S-S |
| 19 | 281.4 | $1.14 \times 10^{-5}$ | $E'$ | S-S |
| 20 | 246.6 | $2.72 \times 10^{-33}$ | $A_1'$ | S-S |

**Table 2.** The calculated electronic transitions of $Bi_5^{3+}$ polycation as shown in Fig. 10b. S-T and S-S represent the singlet-singlet and singlet-triplet transitions, respectively.

| No | Wavelength (nm) | Oscillator strength | Symmetry | type |
|---|---|---|---|---|
| 1 | 1252.4 | 0 | $A_1$ | S-T |
| 2 | 927.9 | 0 | $E$ | S-T |
| 3 | 752.2 | 0 | $A_2$ | S-T |
| 4 | 613.2 | 0 | $E$ | S-T |
| 5 | 549.4 | 0 | $A_1$ | S-T |
| 6 | 478.7 | $3.07 \times 10^{-32}$ | $A_2$ | S-S |
| 7 | 478.4 | 0 | $E$ | S-T |
| 8 | 467.1 | 0 | $E$ | S-T |
| 9 | 466.4 | 0 | $E$ | S-T |
| 10 | 441.8 | $3.60 \times 10^{-3}$ | $E$ | S-S |
| 11 | 434.5 | $2.54 \times 10^{-33}$ | $A_2$ | S-S |
| 12 | 425.8 | $3.91 \times 10^{-4}$ | $E$ | S-S |
| 13 | 390.3 | $9.57 \times 10^{-5}$ | $A_1$ | S-S |
| 14 | 365.7 | $3.99 \times 10^{-6}$ | $E$ | S-S |
| 15 | 352.1 | 0 | $E$ | S-T |
| 16 | 343.1 | 0 | $E$ | S-T |
| 17 | 317.1 | $9.27 \times 10^{-3}$ | $E$ | S-S |
| 18 | 304.3 | $2.16 \times 10^{-3}$ | $A_1$ | S-S |
| 19 | 278.1 | $1.47 \times 10^{-7}$ | $E$ | S-S |
| 20 | 243.9 | $4.62 \times 10^{-4}$ | $A_1$ | S-S |

As shown in Tables 1 and 2, we totally obtain 20 singlet-singlet and singlet-triplet excitation bands for $Bi_5^{3+}$ through above calculation approach. It is found that $Bi_5^{3+}$ polycations in $Bi_5(AlCl_4)_3$ and $Bi_5(GaCl_4)_3$ demonstrate similar theoretical excitation bands, resulting from their similar geometries (Fig. 9). Interestingly, both species have strong singlet-singlet allowed transitions at shorter wavelengths, which support well the experimentally observed excitation spectra as shown in Fig. 5b and 7b. However, at longer wavelengths (500-1000 nm), there remains a great discrepancy between experimental and calculated values, which was also observed when using other quantum chemistry softwares.[23] It is believed that this unsatisfactory consistence might result from the calculation method used here. Owing to high atomic number of bismuth (Z=83), spin-orbit coupling effect should be taken into consideration, which will greatly influences on orbital energies and allowed/forbidden transitions and makes singlet-triplet excitations accessible (i.e., the transitions become allowed).[17b] However, it is worth to note that imaginary eigenvalue appears if we take spin-orbit coupling rather than scalar relativistic effect into account, when using ADF software for the excitation energies calculation. In the range over 1000 nm, singlet-triplet transitions of $Bi_5^{3+}$ at 1415 nm in $Bi_5(AlCl_4)_3$ and 1252 nm in $Bi_5(GaCl_4)_3$ occur, respectively. Unfortunately, using the present calculation method we can not obtain more excitation bands at longer wavelengths. However, in combination with the single-photon excitation-emission characteristics of $Bi_5^{3+}$,[17a] the experimentally observed PL should result from the electronic transition from the excited levels to the ground level. Clearly, more systematical calculation on the UV-Vis-NIR excitation behaviors of $Bi_5^{3+}$ is needed, which may become possible when using other state-of-the-art quantum-chemistry softwares.

## Conclusions

In summary, the photophysical properties of $Bi_5(GaCl_4)_3$ crystal synthesized through the oxidation of Bi metal by gallium chloride salt have been studied experimentally and theoretically. XRD as well as micro-Raman scattering results revealed that the product consists of crystalline $Bi_5(GaCl_4)_3$ and amorphous phase. The thorough comparison of steady-state and time-resolved PL behaviors between $Bi_5(GaCl_4)_3$ product and $Bi_5(AlCl_4)_3$ leads us to conclude that $Bi_5^{3+}$ is the dominant emitter in the product, which gives rise to the ultrabroad emission ranging from 1 to 2.7 μm. Furthermore, we rationalized the experimental results through quantum chemistry calculations. Our work reported here unambiguously indicates that $Bi_5^{3+}$ polycation can be exploited as a smart NIR emitter, and the materials containing it hold great promise for ultrabroad and tunable laser media. Our results demonstrated here, in conjuction with a recent finding on the observation of NIR emission from crystalline (K-crypt)$_2$Bi$_2$ containing $[Bi_2]^{2-}$,[24] suggest that systematical investigation of structural and luminescent properties of molecular crystals containing positively- or negatively-charged bismuth polyhedra allows us to obtain a much clearer picture on bismuth-related photophysical behaviors, because of the establishment of structure-property relationships.[12c,16-18,24] The material systems as well as the method demonstrated here not only serve to solve the confusions on the PL origin of Bi in other material systems including glasses, fibers, and optical crystals,[7-15,19] but also is helpful to develop new applicable laser materials.

## Acknowledgements

H. Sun gratefully acknowledges the funding support from Hokkaido University and NIMS, Japan. H. Sun greatly thanks the fruitful disscussion with Prof. Lars Kloo in Royal Institute of Technology, Sweden, on the synthesis of the sample.

**Notes and references**

[a] *Division of Materials Science and Engineering, Faculty of Engineering, Hokkaido University, Kita 13, Nishi 8, Kita-ku, Sapporo 060-8628, Japan. Fax: +81-11-706-7881; E-mail: timothyhsun@gmail.com*

[b] *International Center for Young Scientists (ICYS), National Institute for Material Sciences (NIMS), 1-2-1 Sengen, Tsukuba-city, Ibaraki 305-0047, Japan.*

[c] *State Key Laboratory of Silicon Materials, Zhejiang University, Hangzhou, Zhejiang, 310027, P. R. China*

[d] *Advanced Ceramics Group, Advanced Materials Processing Unit, National Institute for Materials Science (NIMS), 1-2-1 Sengen, Tsukuba-city, Ibaraki 305-0047, Japan*

[e] *World Premier International Research Center Initiative for Materials Nanoarchitronics (MANA), NIMS, 1-1 Namiki, Tsukuba, Ibaraki 305-0044, Japan*

[f] *PRESTO, Japan Science and Technology Agency (JST), 4-1-8 Honcho Kawaguchi, Saitama 332-0012, Japan*

[g] *Department of Electrical and Electronic Engineering, Kobe University, Kobe 657-8501, Japan*